\documentclass[twocolumn,english]{article}
\usepackage[T1]{fontenc}
\usepackage[latin9]{inputenc}
\usepackage{babel}
\usepackage{cite}
\usepackage{verbatim}
\usepackage{graphicx}
\usepackage[unicode=true, pdfusetitle,
 bookmarks=true,bookmarksnumbered=false,bookmarksopen=false,
 breaklinks=false,pdfborder={0 0 1},backref=false,colorlinks=false]
 {hyperref}
\usepackage{color}
\usepackage{url}
\usepackage{subfig}

\providecommand{\tabularnewline}{\\}

\begin{document}

\title{Back To The Future: On Predicting User Uptime}

\author{Matteo Dell'Amico \and Pietro Michiardi \and Yves Roudier\\
Eurecom, Sophia-Antipolis, France\\
\{matteo.dell-amico,pietro.michiardi,yves.roudier\}@eurecom.fr}

\maketitle
\begin{abstract}
Correlation in user connectivity patterns is generally considered a
problem for system designers, since it results in peaks of demand
and also in the scarcity of resources for peer-to-peer applications. The
other side of the coin is that these connectivity patterns are often
predictable and that, to some extent, they can be dealt with proactively.

In this work, we build predictors aiming to determine the probability
that any given user will be online at any given time in the future. We
evaluate the quality of these predictors on various large traces from
instant messaging and file sharing applications.

We also illustrate how availability prediction can be applied to
enhance the behavior of peer-to-peer applications: we show through
simulation how data availability is substantially increased in a
distributed hash table simply by adjusting data placement policies
according to peer availability prediction and without requiring any
additional storage from any peer.

\end{abstract}

\section{Introduction}\label{sec:introduction}
User uptime patterns in Internet applications are known to be very 
different from what would be obtained from random, uncorrelated 
models. Many measurements 
\cite{chu2002availability,bhagwan2003availability,mickens2006exploiting,guha2006skype,steiner2007kad,javadi-et-al-setiathome-09} 
confirmed that traces of different applications have daily and weekly
patterns. User uptime therefore cannot be modeled as a 
simple Markovian process, because user activity is often correlated. 

In system design, correlation is often seen as a problem: for instance,
simultaneous requests from many users results in a ``flash-crowd'' 
phenomenon which is problematic for content distribution systems; 
in peer-to-peer storage systems, the fact that many user are offline
at the same time creates problems with respect to data availability. 

In this work, we strive to build a set of predictors to exploit the
correlated nature of user activity. Indeed, if users do not
behave randomly, then it should be possible to design mechanisms
capable of anticipating user behavior with a certain degree of
precision. 

A considerable amount of effort has been devoted to characterizing and
predicting session lengths and future uptime patterns within a short
time span \cite{mickens2006exploiting, stutzbach2006churn,
guha2006skype, steiner2007kad, kondo2008correlated}; however,
long-term predictions have been largely neglected, and the
probability for a user to be online is generally modeled as the same for each
user and each moment in the future.

In \cite{mickens2006exploiting}, which is the closest to our work,
uptime predictors are built around the concept of saturating counters
and refinements thereof, and go beyond a boolean classification of
user online time. However, such techniques are not easily amenable to
anticipate the long term user behavior and do not account for users
that abandon an application. 

In this work, we build refined mechanisms for predicting long-term
user behavior that also account for user departures. We verify the
quality of our techniques on traces of Internet applications such as
instant messaging and peer-to-peer applications, and we show that
elaborate predictors are able to consistently reduce the uncertainty
about future user behavior.

Our techniques can be used in many cases where individual user behavior
has an influence on application performance like for example
social networks or peer-to-peer storage applications. To illustrate
the benefits derived from using the information provided by
our predictors, we simulate a distributed hash table (DHT) and show
that an informed policy for choosing node identifiers can result in
higher data availability without requiring additional storage
resources from nodes nor major modifications to the base DHT
mechanism.

\section{Datasets}\label{sec:datasets}
\label{sec:datasets}

In the context of Internet applications, a user generally launches
an application (e.g., a P2P client), establishes a connection to
other users or to a server, and finally disconnects
from the service. We term this series of actions the user's online
behavior. The online behavior is used to compute the \textit{user 
availability}, defined as the cumulative amount of time spent
online, in a reference period of 24 hours.

We analyzed a variety of application traces to study the online
behavior of users and to compute user availability distributions. We
considered an instant messaging application (labelled \textit{IM} in the following),
the eMule file-sharing application relying on the Kad
network \cite{maymounkov2002kademlia} (labelled \textit{Kad}) and the
Skype VoIP application (labelled
\textit{Skype}). For IM, an author of this work is one of the
administrators of a large IM service in Italy and had access to server
logs indicating the online behavior of users. For Kad, we used the traces
collected in \cite{steiner2007kad} and for Skype, we used the dataset
from
\cite{guha2006skype}, obtained by crawling the Skype super-peer network and made available on \cite{kondo2010failure}.
Table~\ref{tab:dataset} summarizes the salient features of the three
datasets: the trace duration ranges from roughly 1 to 6 months and the
number of captured users ranges from roughly 2000 up to several hundred
thousand users.

\begin{table}
\begin{centering}
\begin{tabular}{|c|c|c|c|}
\hline 
Trace & Duration & Users & High availability ($\geq 0.17$)\tabularnewline
\hline
\hline 
IM & 172 days & 1,825 & 354 (19.4\%)\tabularnewline
\hline 
Kad & 179 days & 400,375 & 10,279 (2.57\%)\tabularnewline
\hline 
Skype & 24 days & 2,081 & 1,174 (56.52\%)\tabularnewline
\hline
\end{tabular}
\par\end{centering}
\caption{\label{tab:dataset} Basic dataset information.}
\end{table}

\begin{figure*}[htbp]
\begin{centering}
\begin{tabular}{ccc}
\subfloat[\label{fig:im}IM.]
{\begin{centering}
\includegraphics[width=0.3\textwidth]{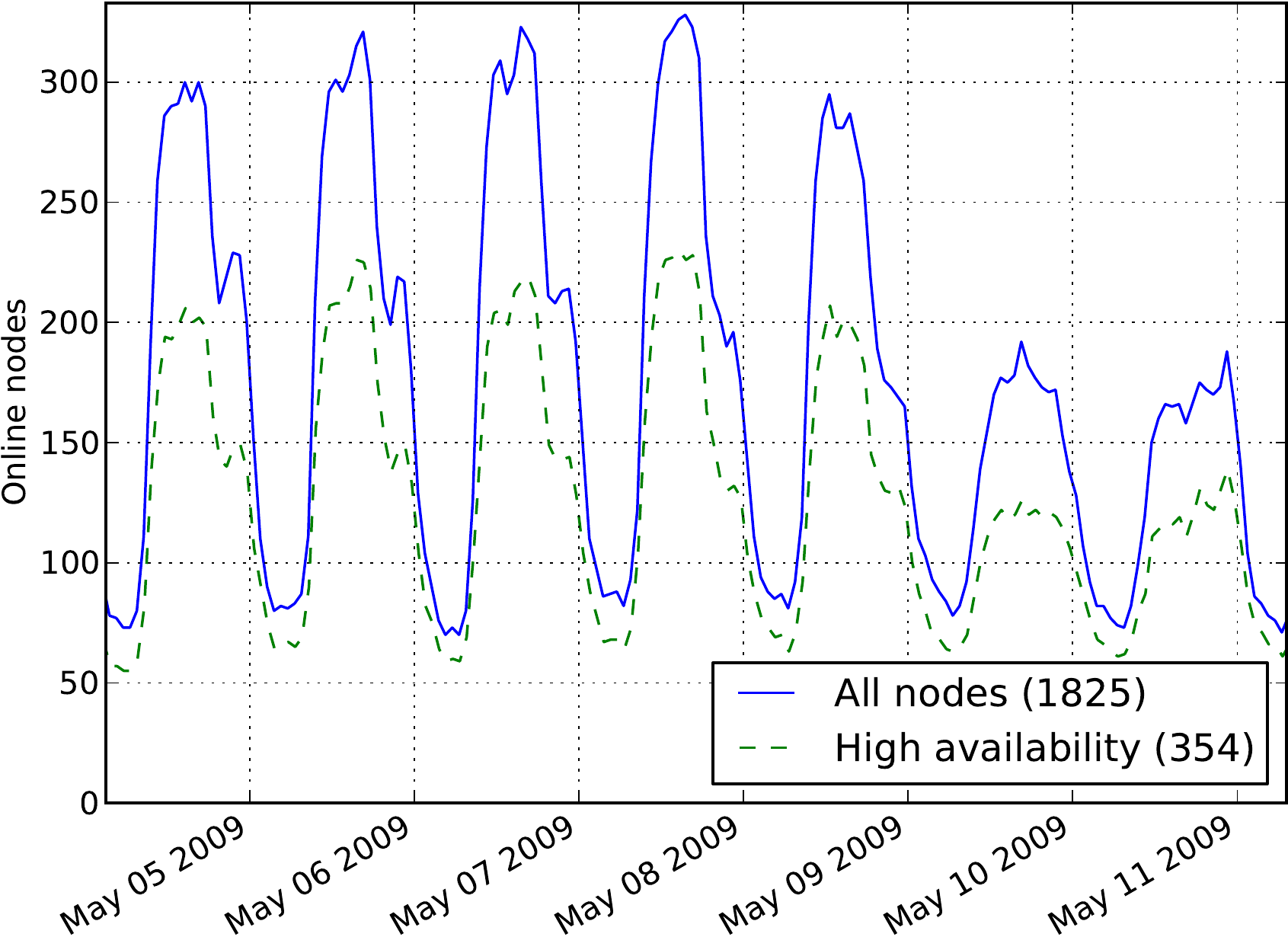}
\par\end{centering}

} & \subfloat[\label{fig:kad}Kad.]{\begin{centering}
\includegraphics[width=0.3\textwidth]{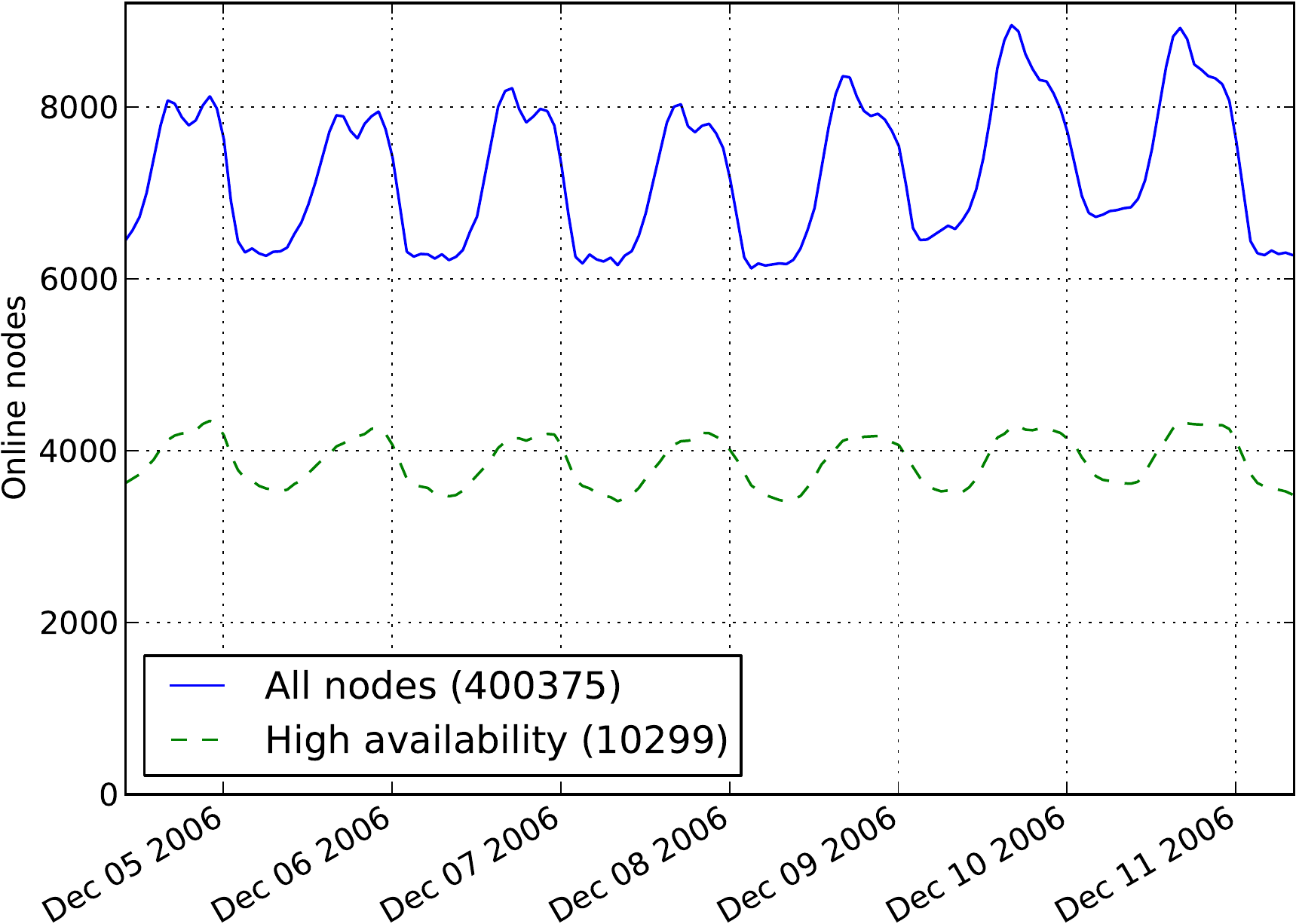}
\par\end{centering}

} & \subfloat[\label{fig:skype}Skype.]{\begin{centering}
\includegraphics[width=0.3\textwidth]{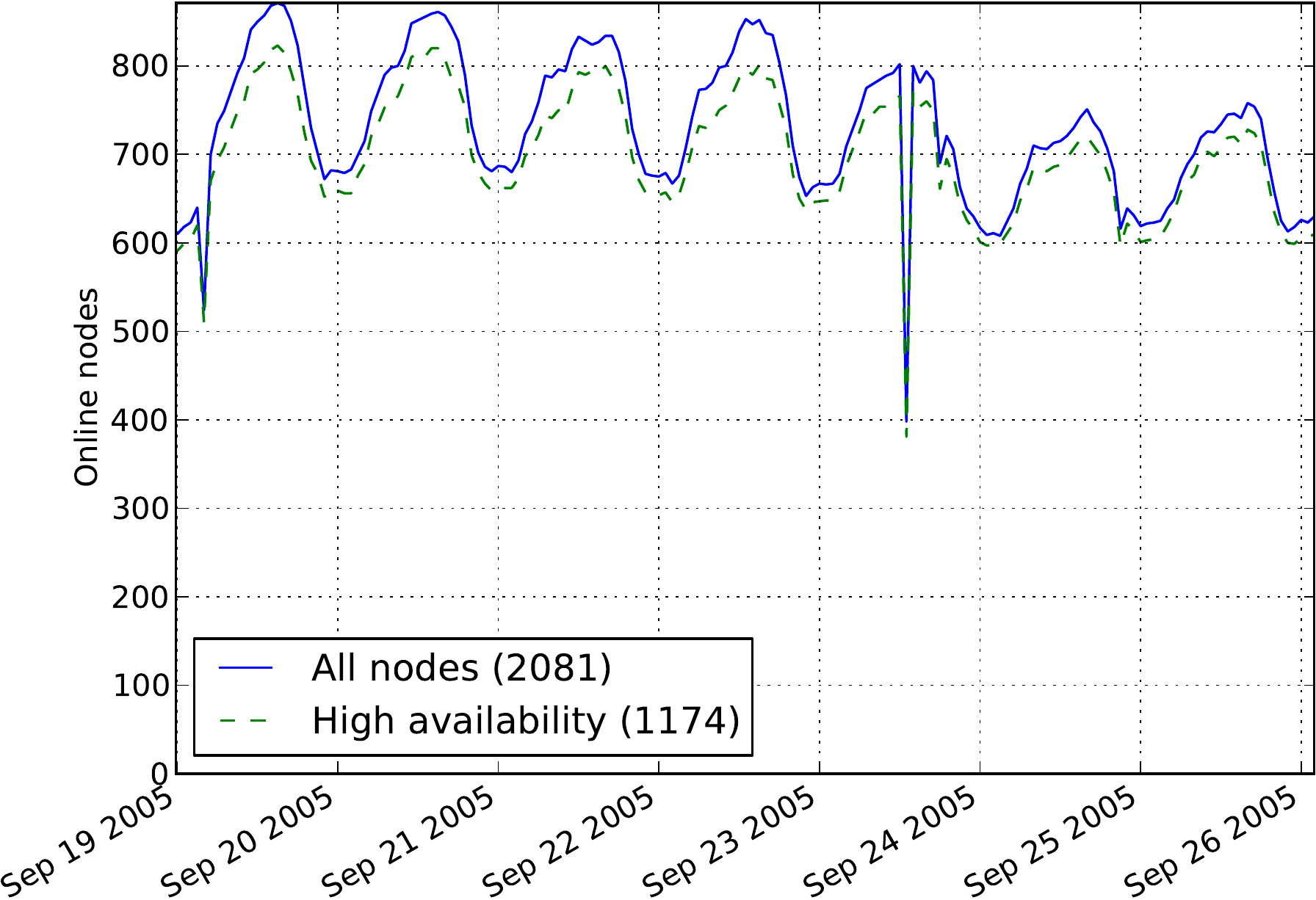}
\par\end{centering}

}\tabularnewline
\end{tabular}
\par\end{centering}

\caption{\label{fig:uptime}Number of online users in an arbitrary week of our different
  datasets.}
\end{figure*}

What information do the above traces convey regarding the online
behavior of users?  Fig.~\ref{fig:uptime} illustrates, for an
arbitrary week of each datasets, the number of online users per day,
detailing users with an availability larger than an average of four
hours per day.  The user behavior is highly correlated: hourly, daily,
and weekly patterns clearly arise. Furthermore, we can pinpoint at
important differences of such patterns depending on the application
examined. In the IM trace, the online behavior is affected by
weekends: in the last two days of the week displayed in
Fig.~\ref{fig:im}, a considerable fraction of users remained
offline. In contrast, the Kad trace indicates a stable online behavior
over a week: users connect mostly at night, which is particularly true
for highly available users.  Clearly, a regularity in the aggregate
traces does not however imply that individual user behavior is
regular.  Lastly, in the Skype trace one can notice that most of the
online users are highly available: this is a result of the crawling
methodology used in
\cite{guha2006skype} which only collects traces of super-peers. Some visible measurement artifacts are due to
network problems on the measurement site.

\begin{figure}
\begin{centering}
\includegraphics[width=0.6\columnwidth]{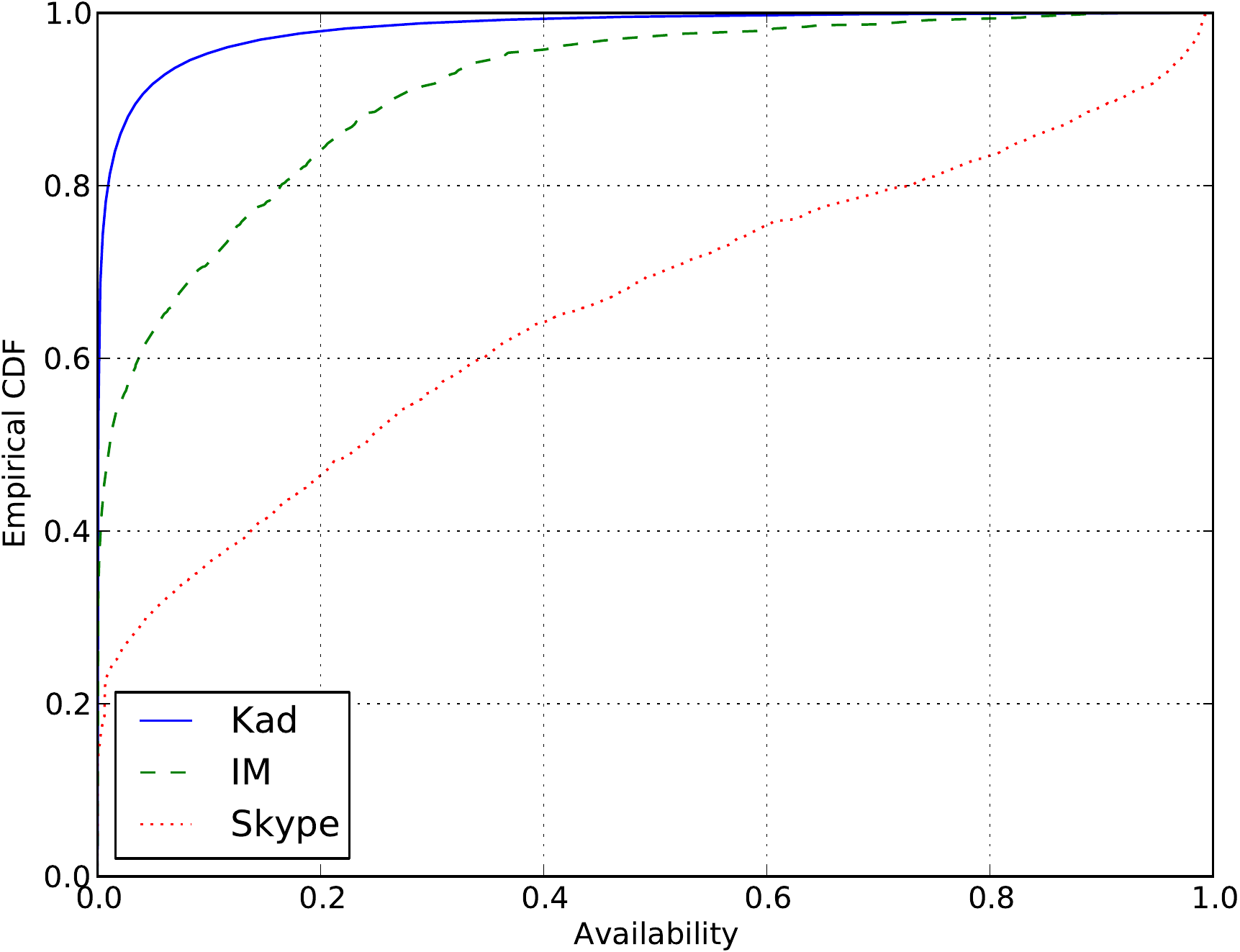}
\par\end{centering}
\caption{CDF of user availability.}
\label{fig:availability}

\end{figure}

The cumulative distribution of user availability is also
clearly distinct for every application trace, as shown
in Fig.~\ref{fig:availability}. Indeed, user availability derives from the
online behavior, as a result of \textit{implicit} or
\textit{explicit incentive mechanisms}.   
%Yves: would there be any reference for such incentive mechanisms in the litterature, especially explicit ones? (even if they are described below ...) [not so important]

%

For the IM application, incentives for users to stay online are
\textit{implicit} and intrinsic to the application
itself. Indeed, IM applications are synchronous, although tolerant to
delays, and require parties to be online at the same time to
communicate. The CDF of user availability (Fig.~\ref{fig:availability})
indicates that a large fraction of the users are sporadically online,
and a small fraction of users have an availability larger than 0.4.

For the Kad application, incentives for users to stay online are
\textit{explicit}. Kad is used to support eMule, a file-sharing
application, which implements a quite elaborate incentive mechanism
that prioritizes users with a high availability when awarding upload
slots \cite{caviglione2008emule}. The CDF of user availability is even
more skewed (Fig.~\ref{fig:availability}), indicating that a very large
fraction of users\footnote{To be precise, we can only characterize
those users that use Kad in combination with eMule, and not all eMule
users.}  are rarely available, while a tiny set of users have an
availability larger than 0.2.

Finally, for the Skype application, incentives are
\textit{implicit}. VoIP applications are not delay tolerant and users
need to be online to be reached by others. 
The distribution of user availability is more uniform than in the
other cases (Fig.~\ref{fig:availability}), apart from an appreciably
small fraction of users that are not available. 

As clearly highlighted above, the user behavior is a combination of
personal factors, like for instance the user's willingness to remain online or user time zone,
and external factors, like application specific incentives or connectivity between hosts.
Given the variety of resulting behaviors, the question we
try and address in the following is whether simple predictors of the future availability of a user
can be designed and tuned, and whether their prediction accuracy is
influenced by the very nature of the application itself.  

Before describing the details of
our prediction techniques, some further observations have to be drawn.
Any attempt at anticipating the online behavior of users would be
doomed to introduce errors if the eventuality for a user to abandon
indefinitely an application was omitted. For this reason, we analyzed the user
\textit{mortality rate} in our traces, defined as the rate of users
``disappearing'' from a dataset. 

\begin{figure}
\begin{centering}
\begin{tabular}{cc}
\subfloat[\label{fig:clustered_traces_IM} IM.]{\begin{centering}
\includegraphics[width=0.45\columnwidth]{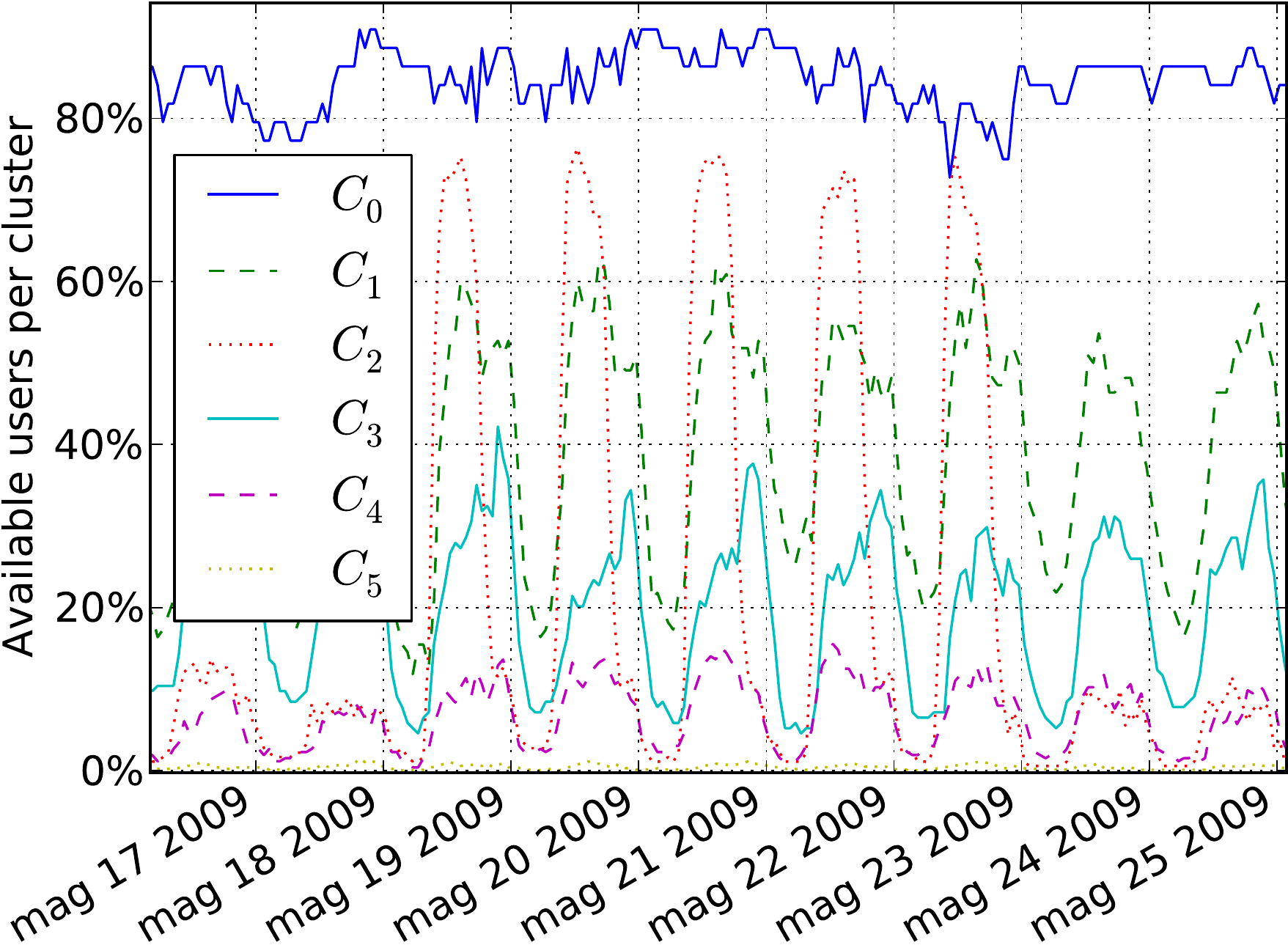}
\par\end{centering}
} & \subfloat[\label{fig:clustered_traces_KAD}Kad. %(10,000 nodes sample)
]{\begin{centering}
\includegraphics[width=0.45\columnwidth]{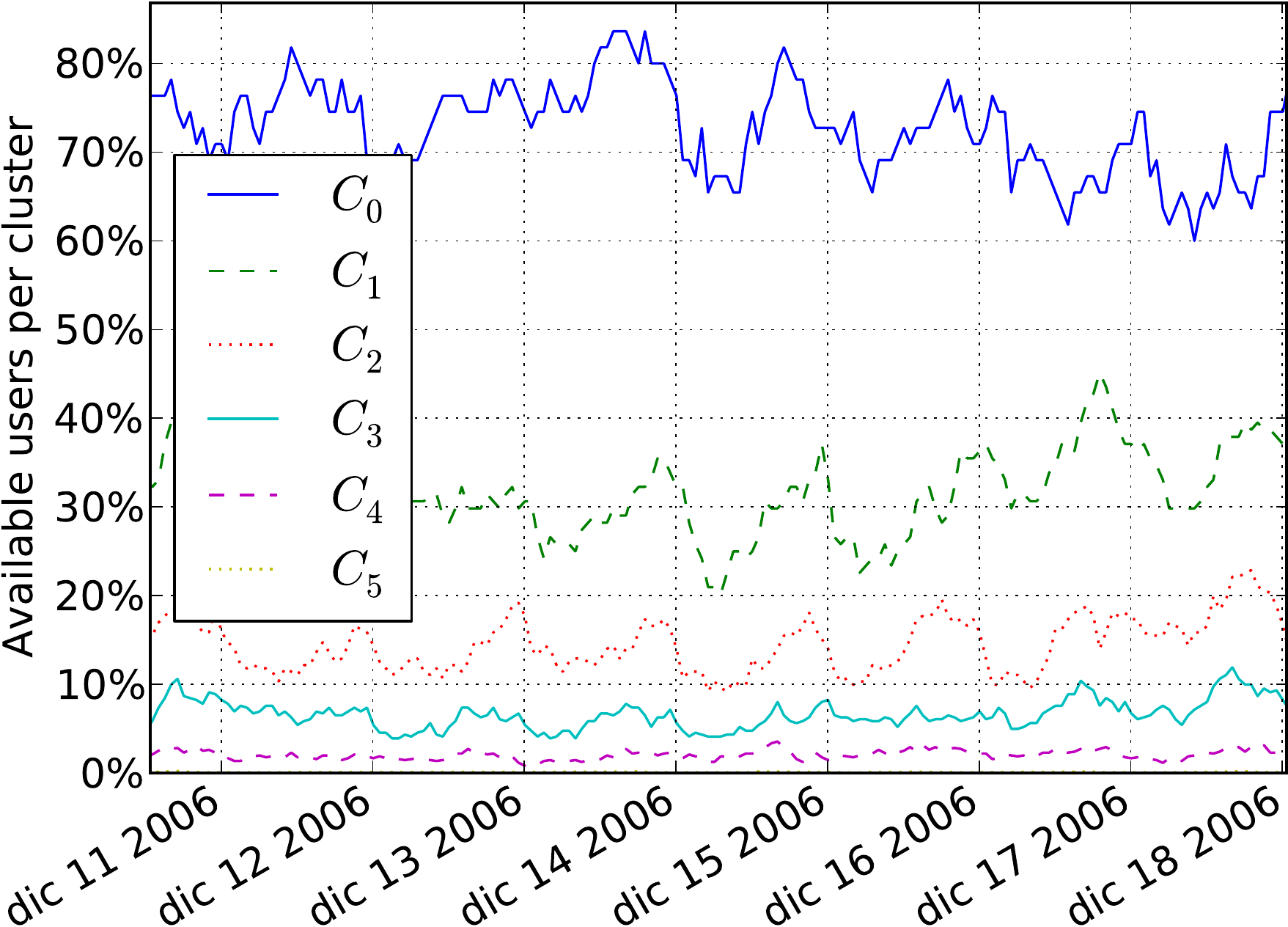}
\par\end{centering}
}\tabularnewline
\end{tabular}
\par\end{centering}
\caption{Clustered traces: detail of single weeks.}
\label{fig:clustered_traces}
\end{figure}

As a second observation, even though most related work focused
on continuous availability estimates, correlated behaviors seem to
be the most critical parameter that needs to be estimated. However,
such correlated behaviors lead to the need for sophisticated predictors
tailored to users rather than attempting to be generic.  In
Fig.~\ref{fig:clustered_traces}, we focus on the IM and Kad
traces %\footnote{For sake of clarity, we focus on a 7 days period of
%the trace.}
and rearrange them by applying an off-the-shelf clustering algorithm
(\emph{$k$-means}). We arbitrarily define $k=6$ clusters (labelled
$C_i$ in the figure) and plot the percentage of online users per
cluster. It can be observed that there are two classes of peers (the first and
the last cluster) that comprise a non-negligible fraction of the total
user population, for which user availability is very high and very
low, respectively. For such users, predicting their availability is
simple. Instead, a large fraction of the user population exhibits very
specific traits. For example, in Fig.~\ref{fig:clustered_traces_IM},
users in $C_1$ have a regular online behavior that is marginally
affected by a particular day of the week, while users in $C_2$ are
highly influenced by weekends, \textit{i.e.,} the last two days displayed in the plot.
Fig.~\ref{fig:clustered_traces_KAD} illustrates another kind of
user-specific behavior: each cluster groups users with consistently
distinct availability figures.

Both observations support our claim that the design of prediction
algorithms, and in particular the tuning phase, should be tailored to
the specific traits of a particular user. An evaluation of the accuracy
of general predictors versus that obtained by individual predictors
is provided in Sec.~\ref{sec:accuracy}. The predictors described
in the following are also adjusted to account for permanent
user departures.

\section{Prediction Algorithms}
\label{sec:Prediction-Algorithms}
To describe the long-term behavior of users, our prediction algorithms
face the task of anticipating, based on a history of past actions, \textit{the
probability} $p_{i,t}$ that a user $i$ will be online at any time $t$
in the future. 
To do so, we divide our traces between a \emph{training period} from
which past observations are drawn and a \emph{test period} in
which predictions are evaluated.  

\begin{figure}
\begin{centering}
\includegraphics[width=0.6\columnwidth]{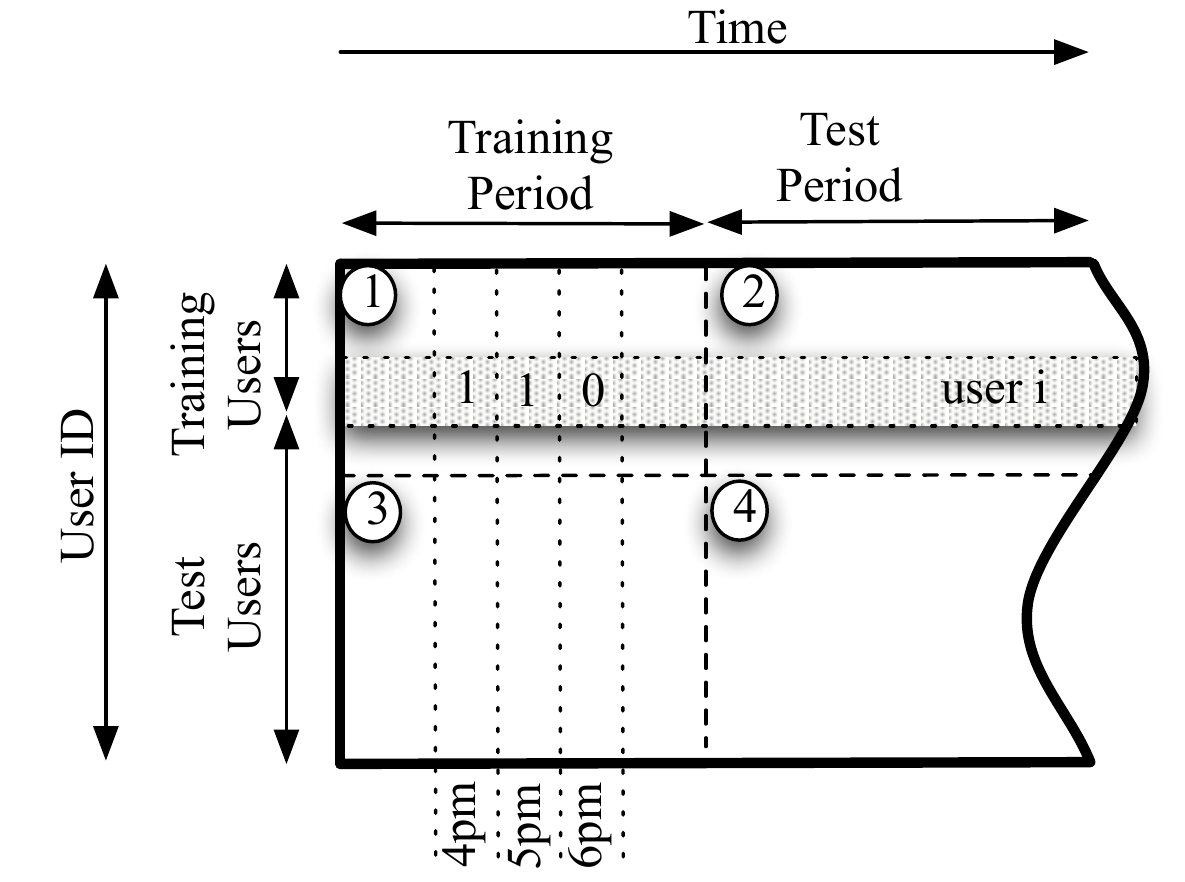}
\par\end{centering}

\caption{Chopping the traces: an illustrative example of our
  four-step approach.}
\label{fig:chopping}

\end{figure}

Since tuning a handful of parameters was required to make our
algorithms work properly, we adopted a four-step approach to the
evaluation, as exemplified in Fig.~\ref{fig:chopping}. Besides
dividing traces between a training and a test period, we distinguished
between the nodes used to ``train'' the algorithm and those used to
validate it.  Predictors are first trained on the first quadrant; in
the ``fitting'' phase, predictors are tuned to provide optimal
performance on the training users for the test period (second
quadrant). In the third phase the predictors, now properly tuned, get
trained with the test users in the training period, and their accuracy
is evaluated on the test periods in the fourth phase.  In a real
situation with growing traces, tuning would naturally be a dynamic
process that would be re-evaluated as the mass of available
information grows.

In the following, we use \emph{Mean Squared Error} (MSE) as a metric
to assess prediction accuracy, considering the prediction error as
$\left(1-p_{i,t}\right)^{2}$ if user $i$ is actually observed online
at time $t$, and $\left(p_{i,t}\right)^{2}$ if $i$ is instead
offline. A ``completely uninformed'' predictor always predicting
$p_{i,t}=0.5$ for any $i$ and $t$ would obtain a MSE of 0.25. The MSE
exhibits a key property (as opposed to other metrics such as, for
example, Mean Absolute Error): if an event has probability
$\overline{p}$, the prediction $p$ that minimizes the MSE is exactly
$p=\overline{p}$. Indeed, the expected MSE is
$\overline{p}\left(1-p\right)^{2}+\left(1-\overline{p}\right)p^{2}=p^{2}-2p\overline{p}+\overline{p}$,
whose differentiation leads to $2\left(p-\overline{p}\right)$. The
function is therefore minimized when $p=\overline{p}$.

We now describe a range of algorithms that differ in the way the
history of past user behavior is processed. For simplicity, we use
Fig.~\ref{fig:chopping} and illustrate the first training phase
described above.  The input to our algorithms is a matrix, where rows
are user identifiers and columns correspond to a moment identified by
time and day of week (e.g., monday, 6PM). The difference in time
between columns is a tunable parameter (1 hour in the example).  Each
cell contains a value indicating the time ratio a user was online at
that particular time during the training period.
Fig.~\ref{fig:chopping} provides example values for user $i$.

In the \textbf{Flat} predictor, we compute the average user
availability for all users and all time-slots, \textit{i.e.}, using
the entire matrix described above. Hence, $p_{i,t}$ equals the average
user availability for all future time instants.
A more refined approach takes into account weekly patterns. In the
\textbf{Weekly periodic} predictor, $p_{i,t}$ is the average
availability of all users for a reference day and time of the week,
that is $p_{i,t}$ is computed for the column identified by the day of
week and hour of $t$. For example, if $t$ corresponds to 6PM on a
Monday, $p_{n,t}$ is the average availability for every Monday at 6PM
in the training period.  The \textbf{Daily periodic} predictor focuses
on daily patterns: hence, $p_{i,t}$ is the average availability of all
users for a reference time of the day, that is $p_{i,t}$ is computed
for all the columns identified by the hour of $t$, irrespectively of
the day of the week.  As users exhibit very different behavior during
the week and during weekends, as illustrated in
Fig.~\ref{fig:clustered_traces}, we designed a \textbf{Weekend-aware
daily periodic} predictor, which isolates weekends from weekdays. This
means that predictions in weekends (resp. weekdays) are only
influenced by observations in weekends (resp. ordinary workdays).

Each approach elaborated above is implemented in two different
``flavors''. A \emph{global} version computes the statistic on all users,
resulting in the same value of $p_{i,t}$ for each user; an
\emph{individual} variant only uses the behavior of user $i$ in the
training period to compute $p_{i,t}$.  

Moreover, we enhance the quality of prediction of all our approaches
as follows (fitting step).  First, we encode the possibility for users
to leave indefinitely the system. We compute the user mortality rate
$r$, as defined in Section~\ref{sec:datasets}, on the training users,
and we update our original prediction to output
$p'_{n,t}=p_{n,t} \cdot r^t$: in our traces, we observed that highly
available users quit an application with a roughly uniform
probability.  Secondly, we compute a linear regression such that the
choice of $a$ and $b$ minimizes the MSE of $p''_{n,t}=ap'_{n,t}+b$ on
the training users, justified by the fact that we in general expect
linear correlation between $p'_{n,t}$ and the actual observations. We
then use $p''_{n,t}$ adjusted with the new values of $r$, $a$, and $b$
as our predictor in the evaluation step.

Note that each of the predictors described so far specializes in
capturing only a single trend of user behavior. A better predictor can
take into account all these factors in order to output a more refined
prediction. Our take at this task is a linear combination of all the
previously defined predictors $p[i]$ (before linear correction): the
resulting predictor is ${\hat p}_{n,t}=\sum_i c_i p[i]_{n,t}$, where
the $c_i$ values are obtained via least-square fitting in order to
minimize errors on the training users.  We call this predictor
\textbf{ad-hoc}, since the values of $c_i$ are different for each
dataset and synthesize the regularities in the trace at hand.

\section{Prediction Accuracy}
\label{sec:accuracy}
% Accuracy is measured in terms of MSE achieved by each class
%of predictors when used on the test period of each trace.
% Before
% describing the results, we first specify how we used the various
% datasets to train the predictors.

In this section, we study the impact of the training period length on the
accuracy of our predictors in terms of MSE.
Both the IM and Kad traces are roughly 6 months long: we use the first
three months of the trace as a candidate training period, while the
test period begins on the first day of the fourth month. 
We therefore considered week, month, and three month long training periods, going
backwards in time from the beginning of the test period (refer to
Fig.~\ref{fig:chopping}).

For the accuracy analysis we filtered all users with an availability
less or equal to 0.17 in the training period\footnote{We use the same
value that the Wuala file storage service adopts to filter peers that
can trade storage~\cite{wuala07}.}: indeed, those are the users whose
behavior is the easiest to predict.
%Yves: this may not be so obvious for the reader or is it? Is it because the traces say so? Why otherwise?
Additionally, for the Kad dataset,
we performed a random sampling of the user population and restricted our
attention to 10,000 training users and 10,000 test users. 
%Yves: do you mean 10,000 and 10,000 or that they are used both for training and test?
The Skype trace is shorter than the other two traces: as a
consequence, we only consider a week-long training period.

%\subsection{Results}
%

\begin{table*}
\begin{center}

\begin{tabular}{|l|l||c|c||c|c||c|c||c|c|c|}

\hline Dataset & Training period &
\multicolumn{2}{c||}{{Flat}} &
\multicolumn{2}{c||}{{Weekly}} &
\multicolumn{2}{c||}{{Daily}} &
\multicolumn{2}{c|}{{``Weekend''}} & {
  Combined}\tabularnewline \cline{3-10} & &
\multicolumn{1}{c|}{{Global}} &
\multicolumn{1}{c||}{{Ind.}} &
\multicolumn{1}{c|}{{Global}} &
\multicolumn{1}{c||}{{Ind.}} &
\multicolumn{1}{c|}{{Glob.}} &
\multicolumn{1}{c||}{{Ind.}} &
\multicolumn{1}{c|}{{Glob.}} &
\multicolumn{1}{c|}{{Ind.}} & {
  ad-hoc}\tabularnewline \hline \hline & 1 week &

           .2037 & .1849 &
            .2036 & .1987 &
            .2037 & .1951 &
            .2034 & .1767 & \textbf{.1727}\tabularnewline
 \cline{2-11} IM &
            1 month & .2039 &
            .1770 & .2036 &
            .1936 & .2037 &
            .1912 & .2032 &
            .1657 & \textbf{.1601}\tabularnewline
 \cline{2-11} & 3 months & .2169 & .1732
            & .2038 & .1933 &
            .2037 & .1877 &
            .2032 & .1517 &
            \textbf{.1478}\tabularnewline \hline \hline
            & 1 week & {.1780} &
                   {.1638} & {.1783} &
                   {.1699} & {.1779} &
                   {.1612} & {.1779} &
                   {.1632} & \textbf{.1608}\tabularnewline \cline{2-11}
                     {Kad} & {1 month} & {.1778} & {.1636} & {.1778} & {.1666} & {.1777} & {.1605} & {.1777} & {.1615} &
                   \textbf{.1598}\tabularnewline
                   \cline{2-11} & {3 months} &
                         {.1779} & {.1707}
                         & {.1780} & {.1697} & {.1779} &
                         {.1664} & {.1779}
                         & {.1671} &
                         \textbf{.1662}\tabularnewline
                         \hline \hline {Skype} &
                                {1 week} &
                                {.2491} & {.2054} & {.2489} &
                                {.2259} & {.2481} & {.1971} &
                                {.2480} & {.2054} & \textbf{.1955}\\

\hline

\end{tabular}
\end{center}

\caption{\label{tab:MSE.}MSE for the various basic predictors (lower is better).}

\end{table*}

Table \ref{tab:MSE.} summarizes the MSE errors for the various
predictors we designed in this work. We report measures for different
training period lengths, as well as for the ad-hoc predictor, which
combines the features of all preceding mechanisms.  It should be
mentioned that comparing the prediction accuracy across the three
dataset reported in this table is somehow irrelevant. For example, the
behavior of Skype users is more difficult to predict than the others,
as the average availability is roughly 0.5. Hence, it is generally
difficult to do better than an uninformed guess of $p_{n,t}=0.5$ that
yields a MSE of 0.25.  Instead, the prediction quality should be
observed within a single dataset, comparing the various predictors to
the Flat predictor.

As a general observation that applies to all our results, it appears
that individual predictors perform better than global ones, which
confirms the intuition that users are characterized by specific
traits, as discussed in Sec.~\ref{sec:datasets}. 
Considering node mortality also ensures consistently
better predictions, especially for the Kad dataset, where user
mortality is higher.

Another global trend that can be observed from Table~\ref{tab:MSE.} is
that prediction accuracy is related to the intrinsic nature of the
datasets we study. For the IM dataset, which involves users connecting
also from work, considering ``specialized'' predictors that include
week days and weekends improves the prediction accuracy. In
comparison, for the Kad dataset, users largely connect from home and
at night and their behavior is not influenced by weekends. Thus,
``specialized'' predictors are not necessarily more accurate.

Finally, the ad-hoc predictor outperforms all other mechanisms we have
designed, confirming that incorporating a range of periodic patterns
effectively increases the prediction quality.

We now discuss the impact of the length of the
training period on prediction accuracy. Global predictors are largely insensitive to training
period lengths: one week of observations on user behavior appears to
be sufficient to reach a plateau for MSE values. Instead, the individual
and ad-hoc predictors are affected by the length of the history of
past user behavior. 
In general, one could think that a longer training period would
mitigate the ``noise'' introduced by a small number of samples on
which the predictors are tuned. However, user behavior
can also evolve with time, and as a consequence, the training phase used to
tune our predictors might use obsolete data.

These observations are verified in our traces.  As the rows
corresponding to the IM trace in Table~\ref{tab:MSE.} show, longer
training periods imply better accuracy, \textit{i.e.,} lower MSE
values. Indeed, the behavior of the users of an IM application is
regular on the long term.  The Kad dataset exhibits an inverted trend:
a longer training period entails lower prediction accuracy. Since the
online behavior of Kad users evolves with time, shorter training
periods are better to reflect these dynamics.

Overall, our results indicate that when properly tuned, our predictors
can effectively anticipate user behavior, as confirmed by 
the low MSE values obtained.
It is of course legitimate to question the concrete meaning of low MSE
values. In particular, what is an acceptable level of accuracy?
Obviously,
it is impossible to design a predictor which makes no errors, and it
is easy to define MSE=0.25 as an upper bound for the prediction
error. %, as this is the error that a random guess would incur.
We try and address this question in the following, where we study the
impact of prediction accuracy in practice, our predictors
being used to optimize the performance of an example application.

\section{An Application Example}
\label{sec:application}
\label{sec:application}
% The predictors described above can be easily adapted to many classes of
% applications like the optimization of data placement in peer-to-peer file storage
% systems \cite{michiardi09selfish} with only small changes.
% We instead select DHTs, which are generic infrastructures mapping
% straightforwardly to the traces we have at hand
% as they can be used in both IM as well as 
% file-sharing applications (as in the case of Kad). 

DHT applications are generic infrastructures mapping straightforwardly
to the traces we have at hand as they can be used in both IM as well
as file-sharing applications (as in the case of Kad). Here, we
consider a Chord-like \cite{stoica2001chord} DHT providing a key-value
lookup primitive.

In our DHT model, identifiers and hash values for keys are distributed
on a logical ring, and each information is replicated on
a \textit{neighbor set} of $n$ nodes whose identifiers are the closest
successors to the hash of the key in the ring. We assume that
information is stored on a long-term basis, so the data does not get
erased from nodes between sessions: hence, data maintenance is
required only when peers abandon the system for good. For simplicity,
we do not implement maintenance mechanisms: data redundancy decreases
with peer ``death''.

In contrast to approaches that reduce object copying in a DHT by
biasing replicas towards highly available nodes
\cite{mickens2006exploiting}, we focus on improving data
availability without imposing additional storage burden on any peer.

In general, node identifiers in DHTs are chosen via a random or
pseudo-random function. We propose instead the application of a smart
policy that maximizes data availability, \textit{i.e.}, the
probability that at least one peer in each neighbor set will be online
at any moment in the future. For example, a smart replica placement
policy would distribute pieces of data between peers which are
frequently online at day and at night in order to obtain high data
availability.
% Yves: I have a problem with the term policy here which seems to
% be used with two different meanings (policy to maximize availability or
% policy to optimize application wrt. peer availability?? I think it is the latter
% so the first sentence might say that the optimization to enhance the probability
% results in a "smart policy", no?)}

The predicted availability of data placed in the DHT can be computed using the ad-hoc predictor
for a neighbor set $N$ and a set of samples in time $T$ as
$$\label{eq:data-availability} 1 - \frac{\sum_{t\in T}1 - \prod_{n\in
    N}(1-p_{n,t})}{|T|}.$$ 
Since our predictors have a weekly period, we limit our analysis to
the first week after the training period, sampling with a frequency of
one hour. 

Our optimizing algorithm works iteratively by repeatedly considering
a pair of random nodes and verifying whether exchanging their
identifiers would enhance, on average, the predicted data availability
for the involved neighbor sets. If so, their identifiers get exchanged. The
algorithm proceeds until swapping operations do not improve data
availability over a fixed threshold.
%Yves: convergence to what? No enhancement above threshold? What if there are two solutions?
 %% {\color{red}(we stop when the
 %%  availability grows by less than $10^{-20}$ in a number of iterations
 %%  comparable to the network size)}. {\color{red}We also implemented
 %%  simulated annealing \cite{} strategies in order to avoid potential
 %%  problems due to local minima, but in practice we didn't obtain real
 %%  benefits from this compared to our simpler default strategy.}
Although centralized in our simulations, this strategy can easily be
implemented in a distributed fashion.
% It is noteworthy that, once the predictors for each node on all
% samples in $T$ are computed, the placement algorithm is easily
% decentralizable: a random DHT lookup is enough to find a random node,
% and the values in Eq. \ref{eq:data-availability} can be obtained based
% on local information.
%Yves: this paragraph is a good candidate to deletion (side note). A
% short sentence with the paragraph above might say: "although centralized
% in our simulations, this strategy can easily be implemented in a distributed
% fashion"

We executed our DHT simulation on the IM and Kad traces with a
training period of 1 month, and on Skype with a training period of 1
week. All results are averaged on 10 simulation runs. Here we compute
the replication factor $n$ using the traditional approach where user
uptime is assumed to be uncorrelated. That is, we used the smallest
$n$ that satisfies $1 - (1 - \overline a)^n\geq 0.99$, where
$\overline a$ is the average availability observed in the training
period.  Applying this formula resulted in a value of $n=15$ for Kad,
$n=11$ for IM, and $n=5$ for Skype. Obviously, our predictions obtain
different values for the estimated data availability, since in
reality user behavior is strongly correlated.

\begin{comment}
\begin{table}

\begin{centering}

\begin{tabular}{|r||c|c||c|c||c|c|}

\hline

Test & \multicolumn{2}{c||}{{Skype}} & \multicolumn{2}{c||}{{IM}} & \multicolumn{2}{c|}{{Kad}} \\

\cline{2-7}

(days) & Rnd. & Opt. &  Rnd. & Opt. &  Rnd. & Opt. \\

\hline \hline

7 & .9833 & .9929 & .9481 & .9788 & .9823 & .9822 \\

\hline

30 & -- & -- & .9509 & .9813 & .9670 & .9796 \\

\hline

60 & -- & -- & .9494 & .9765 & .9439 & .9606 \\

\hline

120 & -- & -- & .9306 & .9580 & .8918 & .9043 \\

\hline

\end{tabular}

\end{centering}

\caption{DHT simulation: availabilities vs. test period.}
\label{tab:dhtsim}

\end{table}
\end{comment}

%% In Table~\ref{tab:dhtsim}, we show the results of sampling the
%% simulated data availability (i.e., the ratio of neighbor sets with at
%% least one online node) with a granularity of one hour and different
%% lengths for the test period. Data availability using the optimized id
%% allocation is consistently better than with a random placement.

%% We also notice some differences between the datasets: IM has a lower
%% data availability on the short run, because of the higher correlation
%% between node availabilities, as apparent from
%% Figure \ref{fig:uptime}. As the training period grows, the average
%% availability decreases because of nodes leaving the system forever;
%% this phenomenon is more noticeable in Kad, due to a higher node
%% turnover rate in the traces. In a realistic application, such a case
%% would be handled by maintenance policies.

\begin{figure}
\begin{centering}
\includegraphics[width=.6\columnwidth]{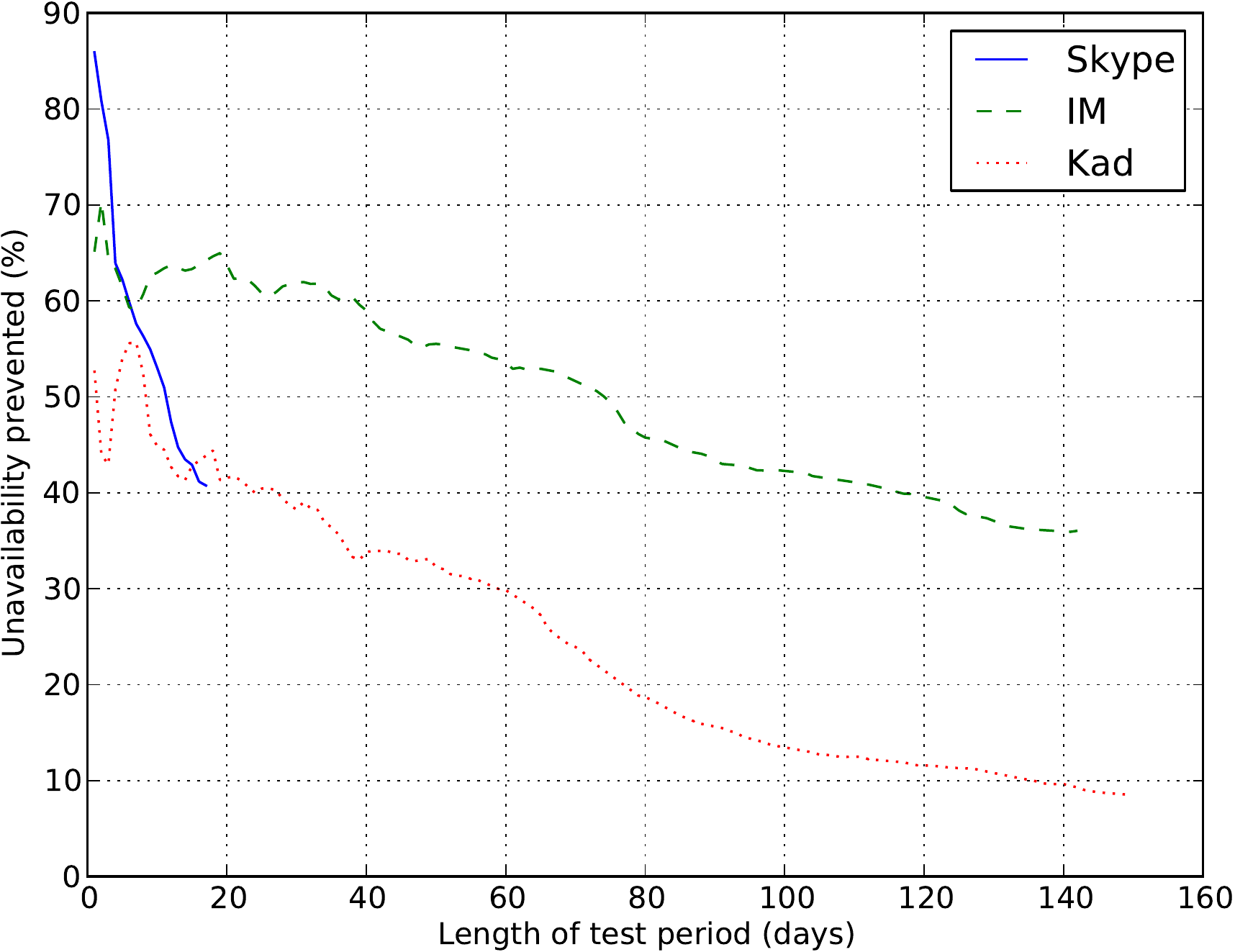}
\par\end{centering}

\caption{\label{fig:dhtsim}DHT simulation: benefits vs. test period.}
\end{figure}

The simulated data availability was computed by sampling the available
nodes in the test period with a granularity of one hour, then
computing the ratio of neighbor sets with at least one online
node. The overall data availability was finally obtained by
considering different lengths for the test period. For example, when a
month is used as a training period, the average simulated data
availability grows in IM from 0.95 to 0.98.

Data availability using the optimized ID allocation is consistently
better than with a random placement. In Fig. \ref{fig:dhtsim} we show
the benefits of the optimized ID allocation in terms of reduced data
unavailability.  For example, a 50\% unavailability reduction means
that the probability that a piece of data is unavailable is halved in
the optimized case with respect to the original random allocation.  As
the test period grows, the benefits of the smart allocation policy
decrease, both because some peers leave the system and because others
change their behavior. In a real system, periodic data maintenance and
identifier reallocation can be used to maintain good performances.
%{\color{red}[PM: note that this does not mean moving data around!]}
%Yves: well moving data around will be necessary for data maintenance!

\section{Conclusions}\label{sec:conclusion}
In this work, we studied the online behavior of users for a range of
Internet applications. We designed and implemented simple predictors
that anticipate user behavior capturing individual, global, daily, and
weekly patterns. We evaluated the accuracy of our mechanisms and
studied their impact on a ``toy'' DHT application, showing that user
behaviors are predictable, which can be used to achieve considerable
benefits in terms of data availability.

We believe that our work can be continued in various interesting
directions. First, better predictors can be designed and tested, in
particular on longer traces once they are available. While there is
obviously an inherent level of unpredictability in the future behavior
of users and even the smartest possible predictor will have a
considerable margin of error, we are at the moment unable to guess if
it is possible to obtain results that are substantially better than
the ones that we are presenting here.

The DHT application that we presented in Section~\ref{sec:application}
is admittedly only a proof of concept. The task of incorporating our
techniques into a real system will incur various tradeoffs,
considering issues such as the cost of running the optimization
algorithm and performing node repositioning. Also, security issues
will need to be examined: could a malicious node be able to exploit
such a repositioning protocol in order to disrupt the system?

Using our predictors to improve data placement in current P2P storage
applications is an important objective. Additionally, we will explore
other applications where availability predictions can be exploited. We
believe that the knowledge of which users will be more likely to
connect at a given moment in time could benefit social networking
applications, e.g., to optimize pre-fetching schemes for home pages of
users which are most likely to connect.

\section*{Acknowledgements}

The authors wish to thank Moritz Steiner and Lluis Pamiez-Juarez for
helping them obtaining the Kad traces.

% Moritz, Lluis

\bibliography{biblio}

\begin{thebibliography}{10}

\bibitem{bhagwan2003availability}
Ranjita Bhagwan, Stefan Savage, and Geoffrey Voelker.
\newblock Understanding availability.
\newblock In {\em Peer-to-Peer Systems II}, pages 256--267, 2003.

\bibitem{caviglione2008emule}
Luca Caviglione, Cristiano Cervellera, Franco Davoli, and Filippo~Aldo Grassia.
\newblock Optimization of an emule-like modifier strategy.
\newblock {\em Computer Communications}, 31(16):3876 -- 3882, 2008.
\newblock Performance Evaluation of Communication Networks (SPECTS 2007).

\bibitem{chu2002availability}
Jacky Chu, Kevin Labonte, and Brian~N. Levine.
\newblock Availability and locality measurements of peer-to-peer file systems.
\newblock In {\em Proc. of ITCom: Scalability and Traffic Control in IP
  Networks}, 2002.

\bibitem{wuala07}
D.~Grolimund.
\newblock Wuala - a distributed file system.
\newblock Google {T}ech{T}alks video,
  \url{http://www.youtube.com/watch?v=3xKZ4KGkQY8}, 2007.

\bibitem{guha2006skype}
Saikat Guha, Neil Daswani, and Ravi Jain.
\newblock An experimental study of the skype peer-to-peer voip system.
\newblock In {\em Proc. IPTPS}, 2006.

\bibitem{javadi-et-al-setiathome-09}
Bahman Javadi, Derrick Kondo, Jean~M. Vincent, and David~P. Anderson.
\newblock {M}ining for {A}vailability {M}odels in {L}arge-{S}cale {D}istributed
  {S}ystems:{A} {C}ase {S}tudy of {SETI}@home.
\newblock In {\em MASCOTS 2009}. IEEE, September 2009.

\bibitem{kondo2008correlated}
D.~Kondo, A.~Andrzejak, and D.~P. Anderson.
\newblock On correlated availability in internet-distributed systems.
\newblock In {\em Proceedings of the 2008 9th IEEE/ACM International Conference
  on Grid Computing}, pages 276--283. IEEE Computer Society, 2008.

\bibitem{kondo2010failure}
D.~Kondo, B.~Javadi, A.~Iosup, and D.~Epema.
\newblock The failure trace archive: Enabling comparative analysis of failures
  in diverse distributed systems.
\newblock In {\em 10th IEEE/ACM International Symposium on Cluster, Cloud and
  Grid Computing (CCGrid)}, 2010.

\bibitem{maymounkov2002kademlia}
P.~Maymounkov and D.~Mazi{\`e}res.
\newblock {Kademlia: A Peer-to-Peer Information System Based on the XOR
  Metric}.
\newblock In {\em Revised Papers from the First International Workshop on
  Peer-to-Peer Systems}, pages 53--65. Springer-Verlag, 2002.

\bibitem{mickens2006exploiting}
J.W. Mickens and B.D. Noble.
\newblock Exploiting availability prediction in distributed systems.
\newblock In {\em Proceedings of the 3rd conference on Networked Systems Design
  \& Implementation-Volume 3}, page~6. USENIX Association, 2006.

\bibitem{steiner2007kad}
Moritz Steiner, Taoufik~E. Najjary, and Ernst~W. Biersack.
\newblock A global view of {K}ad.
\newblock In {\em IMC '07: Proceedings of the 7th ACM SIGCOMM conference on
  Internet measurement}, pages 117--122, New York, NY, USA, 2007. ACM.

\bibitem{stoica2001chord}
I.~Stoica, R.~Morris, D.~Karger, M.F. Kaashoek, and H.~Balakrishnan.
\newblock {Chord: A scalable peer-to-peer lookup service for internet
  applications}.
\newblock In {\em Proceedings of the 2001 conference on Applications,
  technologies, architectures, and protocols for computer communications}, page
  160. ACM, 2001.

\bibitem{stutzbach2006churn}
Daniel Stutzbach and Reza Rejaie.
\newblock Understanding churn in peer-to-peer networks.
\newblock In {\em IMC '06: Proceedings of the 6th ACM SIGCOMM conference on
  Internet measurement}, pages 189--202, New York, NY, USA, 2006. ACM.

\end{thebibliography}
\bibliographystyle{plain}

\end{document}